\documentclass[twocolumn]{article} 

\usepackage{graphicx} 
\usepackage{citesort}
\usepackage[usenames]{color}
\usepackage{amsmath,amssymb}
\usepackage{gensymb}
\begin{document} 

\title{Diffusive shielding stabilizes bulk nanobubble clusters}

\author{Joost H. Weijs$^1$, James R. T. Seddon$^1$, and Detlef Lohse$^{1,2}$}

\date{\today}

\twocolumn[
\maketitle
  \begin{@twocolumnfalse}
    \maketitle
    \begin{abstract}
Using molecular dynamics, we study the nucleation and stability of bulk nanobubble clusters.
We study the formation, growth, and final size of bulk nanobubbles.
We find that, as long as the bubble-bubble interspacing is small enough, bulk nanobubbles are stable against dissolution.
Simple diffusion calculations provide an excellent match with the simulation results, giving insight into the reason for the stability: nanobubbles in a cluster of bulk nanobubbles ``protect'' each other from diffusion by a shielding effect.

\end{abstract} 
  \vspace{1cm}

  \end{@twocolumnfalse}
  ]

\footnotetext[1]{Physics of Fluids Group and J. M. Burgers Centre for Fluid Dynamics,University of Twente, P.O. Box 217, 7500 AE Enschede, The Netherlands}
\footnotetext[2]{d.lohse@utwente.nl}
\section{Introduction}
Gas bubbles are ubiquitous in nature, industry and daily life.
They are found in streams of water, manufacturing processes of many types of materials, and, of course, when we enjoy a carbonated drink.
Even though gas bubbles are commonly present in many of the liquids we deal with on a daily basis, bubbles are, in fact, usually unstable against dissolution in the medium that surrounds them~\cite{Epstein50}.
The dissolution rate increases as the bubble becomes smaller because of the increased (Laplace) pressure $\Delta p=2\gamma/R$ inside the bubble, where $\gamma$ is the interfacial tension of the bubble wall and $R$ the bubble radius.
The consequence is that nanoscopic bubbles cannot survive for more than a few microseconds.

In contrast to this expectation, surprisingly, experiments by Ohgaki~\emph{et al.}~\cite{Ohgaki10} have shown that stable bulk nanobubbles do exist.
In these experiments the bubbles were observed to be packed closely together (the distance between neighbouring bubbles was measured to be less than $10R$), suggesting that a shielding mechanism between bubbles may act to keep the bubbles from dissolving.
In addition to this direct observation of bulk nanobubbles, their presence has also been indirectly measured in experiments, using dynamic laser light scattering~\cite{Jin07,Jin07_2}.
Although this technique cannot distinguish between nanobubbles and liquid density variations in the liquid caused by other sources (such as large organic molecules), the observed fluctuations disappear after degassing the liquid, indicating that the observed objects are indeed bulk nanobubbles.

In addition to these experiments, there are many publications where the presence of surface nanobubbles are observed at liquid-solid interfaces. Generally, these surface nanobubbles are detected by Atomic Force Microscopy (AFM), and they can survive for days \cite{Hampton,Craig11,Seddonrev11}.
Similar to bulk nanobubbles, surface nanobubbles should dissolve within microseconds, in contrast to the AFM observations~\cite{Parker94,Lou2000,Tyrrell01,Holmberg03,Steitz03,Simonsen04,Zhang04,BorkentPRL,Yang07,Yang08,Hampton,Craig11}.
Various stabilization mechanisms have been proposed \cite{Ducker09,Brenner,Hampton,Craig11,Seddonrev11}, and many of them invoke the direct bubble-wall interaction. 
This in particular holds for the dynamic equilibrium theory promoted by some of us~\cite{Brenner,Seddonrev11,SeddonPRL11}.
This stabilization mechanisms is therefore not applicable to bulk nanobubbles: the symmetry breaking caused by the presence of the substrate in the case of surface nanobubbles does not exist for bulk nanobubbles.

On the other hand, different stabilization mechanisms may exist that could account for stable bulk nanobubbles. Such a mechanism will be discussed in this paper: when a bulk nanobubble is surrounded by more nanobubbles, the diffusive outflux is `shielded': a locally high concentration of dissolved gas in the water suppresses the diffusive outflux from the bubble. For this to happen a cluster of bubbles must exist where the spacing between bubbles is not too large.
Indeed, the bulk nanobubbles reported by Ohgaki~\emph{et al.}~\cite{Ohgaki10} have a distance of $10R$ or less.

\begin{figure}[ht]
\begin{center}
  \includegraphics[width=75mm]{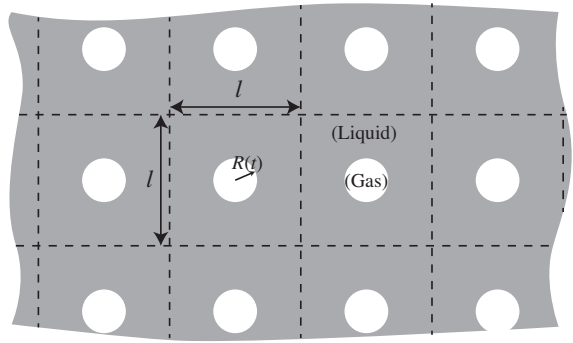}
  \caption{\label{fig:bubcluster}A cluster of nanobubbles on a rectangular grid. The dotted lines indicate the unit cell, which is rectangular and has sides with length $l$: the distance between two neighbouring bubbles.}
\end{center}
\end{figure}

In this paper, we will discuss molecular dynamics (MD) simulations of binary mixtures of simple (Lennard-Jones) fluids. One of the fluids is under the imposed conditions ($T=300$K, $p=10^5$~Pa) in the liquid state, the other in the gaseous state. 
The simulations will be carried out in a simulation domain of which one dimension is very small ($\ell\times\ell\times d, d\ll\ell$), such that the simulations are quasi-2D, see also Fig.~\ref{fig:bubcluster}.
For a full 3D case, the results will only differ quantitatively, but qualitatively they will be the same.
Periodic boundary conditions are applied in all directions, such that we only have to simulate one single nanobubble which is then mirrored.
This infinite repetition of nanobubbles then represents an infinite (periodic) nanobubble cluster ( see Fig.~\ref{fig:bubcluster}) in a closed system. 
The closed system means that the total amount of gas is conserved.
In this work, we explore two box sizes: $\ell = 15$~nm and $\ell = 30$~nm, with $d = 3.64$~nm in both cases. 

The paper is organized as follows: In section~\ref{sec:numdet} the numerical details of the simulations will be outlined, such as the parameters and algorithms used, as well as the initial conditions.
Next, in section~\ref{sec:results}, the results of the MD-simulations will be presented and discussed, and in section~\ref{sec:cont} we will compare these results (particularly the equilibrium radius $R_{eq}$ of bulk nanobubbles) with continuum predictions.
Finally, in section~\ref{sec:cluststab} we will discuss the stability of the entire cluster (as opposed to just single bubbles inside the cluster).

\section{Numerical details}
\label{sec:numdet}
\subsection{Molecular Dynamics}
To simulate a cluster of bulk nanobubbles, we use Molecular Dynamics (MD) simulations of simple fluids.
The atoms in the simple fluids interact with each other through the Lennard-Jones potential:
\begin{equation}
U_{LJ}(r) = 4\epsilon_{ij} \left[ \left( \frac{\sigma_{ij}}{r}\right)^{12}-\left( \frac{\sigma_{ij}}{r}\right)^6 \right]\;.
\end{equation}
Here, $\epsilon_{ij}$ is the interaction strength between atom species $i$ and $j$, and $\sigma_{ij}$ the interaction radius between atoms species $i$ and $j$.
In our simulations we use two atom types: the first is in the liquid state under the conditions considered ($p=10^5$~Pa, $T=300$K) and the second in the gas state.
The interactions are defined as follows: $(\epsilon_{ll},\epsilon_{gg},\epsilon_{lg}) = (3,1,\sqrt{\epsilon_{ll}\epsilon_{gg}}=1.73)$~kJ/mol, $(\sigma_{ll},\sigma_{gg},\sigma_{lg}) = (0.34,0.5,(\sigma_{ll}+\sigma_{gg})/2=0.42)$~nm.
The simulations are carried out in the NPT-ensemble (constant number of particles, pressure, and temperature).
A Berendsen pressure scaling algorithm was applied, and the temperature was kept constant using a velocity rescaling thermostating procedure~\cite{Bussi07}.
\subsection{Initial conditions}
\begin{figure}[ht!]
\begin{center}
  \includegraphics[width=75mm]{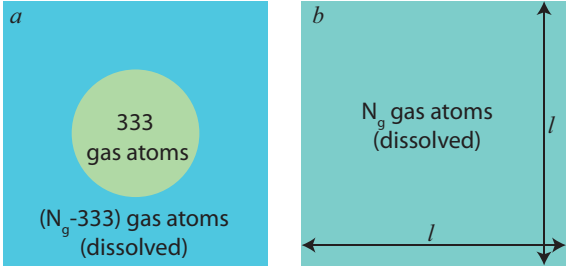}
  \caption{\label{fig:ICs}The two types of initial conditions for the simulations. a) A preformed bubble containing 333 gas-atoms surrounded by liquid. If there are more gas atoms in the system ($N_g>333$) they are uniformly dissolved throughout the liquid. b) All ($N_g$) gas atoms are uniformly dissolved throughout the liquid, so there is no pre-formed bubble. If the concentration of gas is high enough, homogeneous nucleation will occur forming a nanobubble.}
\end{center}
\end{figure}
We use two different atom start-position configurations, which are shown in Fig.~\ref{fig:ICs}.
The first configuration consists of a preformed bubble at a predefined radius $R_0$ containing gas (333 atoms) and vapour.
Outside the bubble the simulation box is completely filled with liquid, and the remainder of the gas is uniformly dissolved throughout the liquid.
For the second configuration, the simulation box is completely filled with liquid with the gas uniformly dissolved in this liquid (so no pre-existing bubble).
In this configuration, a nanobubble will occur if the concentration of gas in the liquid is high enough such that the energy barrier for homogeneous nucleation can be overcome.
Since the pressure is maintained constant throughout the simulation, the box-size is allowed to vary to accommodate this.
In practice, we find that the box dimensions never vary more than 10$\%$ from their initial values.
The initial velocities for all atoms are sampled from a Maxwell-Boltzmann distribution at 300K.
\begin{table*}[th]
\caption{\label{tbl:pars}Simulation parameters of the different simulations. The Initial Conditions (IC) type refers to the configurations shown in Fig.~\ref{fig:ICs}.}
\vspace{.3cm}
\begin{center}
\begin{tabular}{l|c c c c c c c}
Exp. & $\ell$[nm] & $N_g$[$\#$] & $N_l$[$\#$] &$N_g/N_l$& $N_g/\sqrt{N_l}$& IC type (Fig.~\ref{fig:ICs})&Stable?\\ 
\hline\hline I & 15 & 333 & 12339 & 2.70$\cdot 10^{-2}$&3.0&a&yes \\ 
II & 15 & 342 & 12330 &2.77$\cdot 10^{-2}$&3.1&a& yes\\
III & 15 & 432 & 12240 &3.53$\cdot 10^{-2}$&3.9&a& yes\\
IV & 15 & 531 & 12141 & 4.44$\cdot 10^{-2}$&4.8&a& yes\\
V & 30 & 333 & 52489 & 0.63$\cdot 10^{-2}$&1.5&a& no\\
VI & 30 & 342 & 52480 & 0.65$\cdot 10^{-2}$&1.5&a& no\\
VII & 30 & 432 & 52390 & 0.82$\cdot 10^{-2}$&1.9&a& yes\\
VIII & 30 & 832 & 51990 & 1.60$\cdot 10^{-2}$&3.6&a& yes\\
I-b & 15 & 332 & 12340 & 2.69$\cdot 10^{-2}$&3.0&b &yes\\
III-b & 15 & 436 & 12236 & 3.56$\cdot 10^{-2}$&3.9&b&yes\\
V-b & 30 & 333 & 52489 & 0.63$\cdot 10^{-2}$&1.5&b& no nucleation\\
VII-b & 30 & 432 & 52390 & 0.82$\cdot 10^{-2}$&1.9&b& no nucleation\\
VIII-b & 30 & 837 & 51985 & 1.61$\cdot 10^{-2}$&3.7&b& no nucleation
\end{tabular} 
\end{center}
\end{table*}
\section{Results from the MD simulations}
\label{sec:results}
\begin{figure*}[ht!]
\begin{center}
  \includegraphics[width=160mm]{Fig3.eps}
  \caption{\label{fig:snaps}Snapshots from a selection of simulations. Note that the bubbles do not remain centered in the simulation box due to Brownian motion. This is not a problem since periodic boundary conditions are imposed, such that the bubble moves back into the simulation domain at the opposite site from which it leaves. Note also how the box size adapts to keep the pressure in the system constant. \emph{Left}: Two simulations (III~and~VI) with initial conditions type a (Fig.~\ref{fig:ICs}). For simulation III, the bubble grows towards a stable radius. For simulation VI (larger $\ell$), the bubble completely dissolves within $\sim 70$~ns. \emph{Right}: Bubbles simulated using initial conditions type b (Fig.~\ref{fig:ICs}). For simulation~III-b, homogeneous nucleation occurs and the bubble grows towards the same equilibrium radius as the bubble in simulation~III. For simulation~VIII-b, no nucleation occurs and the gas remains homogeneously dissolved throughout the liquid.}
\end{center}
\end{figure*}
\begin{figure*}[ht!]
\begin{center}
  \includegraphics[width=150mm]{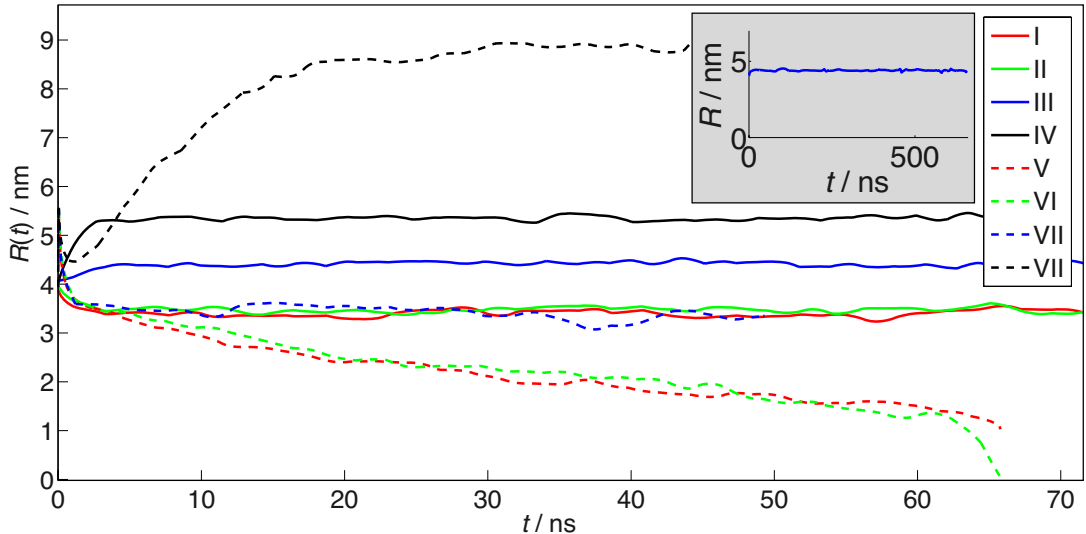}
  \caption{\label{fig:Rtt}Results from the simulations with initial condition type a (Fig.~\ref{fig:ICs}). All bubbles in the small box ($\ell=15$nm) are stable (I-IV), whereas the simulations in the large box ($\ell=30$nm) and with little gas initially dissolved in the liquid (V,VI) are unstable. When a sufficient amount of gas is initially dissolved in the liquid (VII, VIII) the bubbles are stable even in the large box. The inset shows the extended simulation of configuration III, to verify that the bubble is indeed stable at long timescales.}
\end{center}
\end{figure*}

A total of 8 different bubbles have been simulated (see Table~\ref{tbl:pars}) which started with a pre-existing bubble in the initial conditions (Fig.~\ref{fig:ICs}a).
Of those configurations 5 additional simulations were performed using the initial conditions without a pre-existing bubble (Fig.~\ref{fig:ICs}b), to see whether the initial conditions affect the final result.
The boundary of the bubble is defined at $\rho^\ast=0.5$, where:
\begin{equation}
\rho^\ast(\vec{r}) = \frac{\rho(\vec{r})-\rho_v}{\rho_l-\rho_v}\;.
\end{equation}
Here, $\rho_v$ is the bulk number density of the gas/vapour phase inside the bubble and $\rho_l$ the number density in the bulk liquid.
This boundary is then fitted with a circle giving $R(t)$.
Some snapshots of a selection of simulations are shown in Fig.~\ref{fig:snaps}.
$R(t)$ against time is plotted in Fig.~\ref{fig:Rtt} where we see that some bubbles are stable, while others are not.
As one would intuitively expect, the bubbles that are closest together ($\ell=15$nm, configurations I-IV) are stable, whereas some bubbles that are spaced further apart ($\ell=30$nm) are not.
The stable bubbles benefit from their nearest neighbours, as they `shield' the diffusive outflux that would normally lead to dissolution within microseconds.
To confirm that these bubbles are truly stable, we extended one simulation (configuration III) until $t=0.8\;\mu$s, where we found that after $t=3$~ns the radius $R$ remained perfectly constant (see inset Fig.~\ref{fig:Rtt}).

The bubbles that are spaced farther from each other ($\ell = 30$nm, configurations V-VIII) are not always stable.
The configurations with the least amount of gas (configurations V and VI) dissolve within 70~ns, whereas configurations VII and VIII \-- which contain more gas \-- are stable.

\begin{figure}[t]
\begin{center}
 \includegraphics[width=75mm]{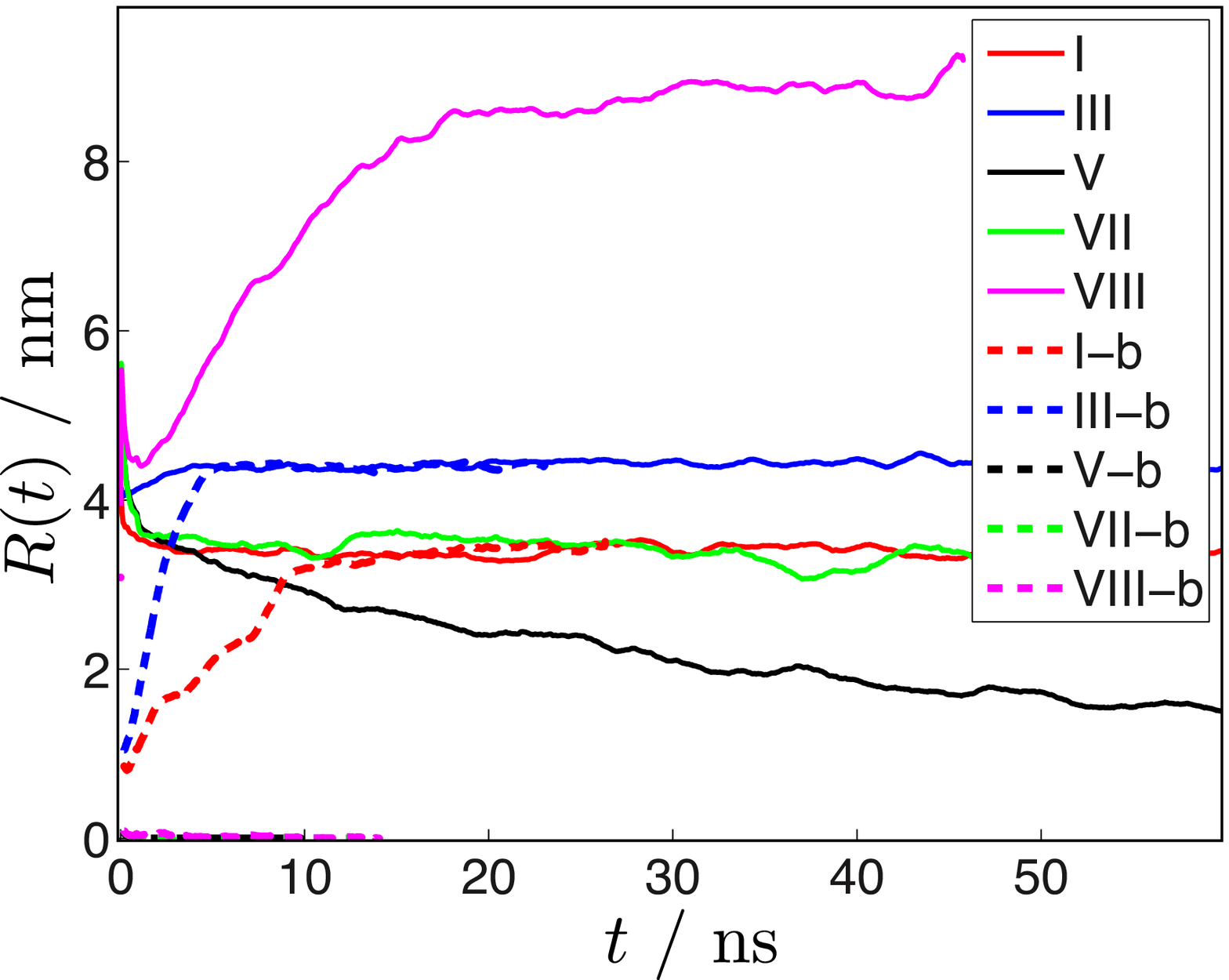}
\caption{\label{fig:altIC}Bubble radius evolution as a function of time comparing similar systems with different initial conditions. When bubbles nucleate, they grow towards the same equilibrium size independent of initial conditions. Although a stable bubble exists for configurations VII and VIII, the bubbles in configurations VII-b and VIII-b do not reach this state as the bubbles do not nucleate. Naturally, nucleation also does not occur for the configuration where no stable bubble can exist (V, V-b). }
\end{center}
\end{figure}
What happens when the initial conditions are changed from a pre-existing bubble (Fig.~\ref{fig:ICs}a) to uniformly dissolved gas in bulk liquid (Fig.~\ref{fig:ICs}b)?
If a bubble forms, there is no reason why it shouldn't grow to the same equilibrium size as the corresponding bubbles with different initial conditions.
In fig.~\ref{fig:altIC} we show the results of some simulations with the alternative initial conditions (I-b, III-b, V-b, VII-b, and VIII-b) compared to the data of the similar bubbles with the original initial conditions (I, III, V, VII, and VIII).
We see that when nanobubbles form, they indeed grow towards the same equilibrium size which they also achieve when starting with a finite size bubble.
The reason that for these initial conditions in some cases no bubbles form is that the gas concentration $\phi$ in the liquid is not large enough to overcome the nucleation barrier.

\section{Continuum description}
\label{sec:cont}
In this section, we will use continuum fluid mechanics to explain and predict the behavior of nanobubble clusters. In particular, we will address their stability and calculate their equilibrium size. We will however first address the subject of homogeneous nucleation, which is relevant for the simulations where there was no pre-existing bubble in the initial conditions.
\subsection{Nucleation theory}
\label{sec:nucleation}
In the case where there is no pre-existing bubble in the initial conditions (Fig.~\ref{fig:ICs}b), we are dealing with homogeneous nucleation, since there are no seeds (such as contamination) available to start heterogeneous nucleation~\cite{Schmelzer}.
In the case of homogeneous nucleation, the change in free energy of the system when a bubble of radius $R$ forms is:	
\begin{equation}
\Delta G = \frac{4}{3}\pi R^3 G_v + 4 \pi R^2 \gamma\;.
\end{equation}
Here, $\Delta G$ is the energy gain or loss for the system to form a bubble of radius $R$.
$G_v$ is the (volumetric) energy associated with a unit volume of gas and is a negative number, hence it promotes nucleation.
The liquid-vapour surface tension $\gamma$ is always positive and therefore acts agains nucleation.
For small $R$, the surface energy term usually wins, but when the bubble reaches a critical radius $R^\ast=-2\gamma/G_v$ then d$G$/d$R|_{R>R^\ast} < 0$, meaning that the bubble will grow.
In our case, we start out with a bubble of zero radius, but due to thermal fluctuations small bubbles appear randomly throughout the system.
By increasing the gas concentration in the liquid, the magnitude of $G_v$ increases as it becomes more and more favourable for the system to have gas atoms in the gas phase than in the dissolved state.
This decreases the value of $R^\ast$, until it is small enough that the spontaneously forming tiny bubbles are already large enough ($R>R^\ast$) to overcome the surface energy penalty and grow.
The growth stops when equilibrium between the gas and liquid phase is achieved [Eq.~\eqref{eq:Henry}].
\subsection{Growth dynamics}
\begin{figure}[ht]
\begin{center}
  \includegraphics[width=80mm]{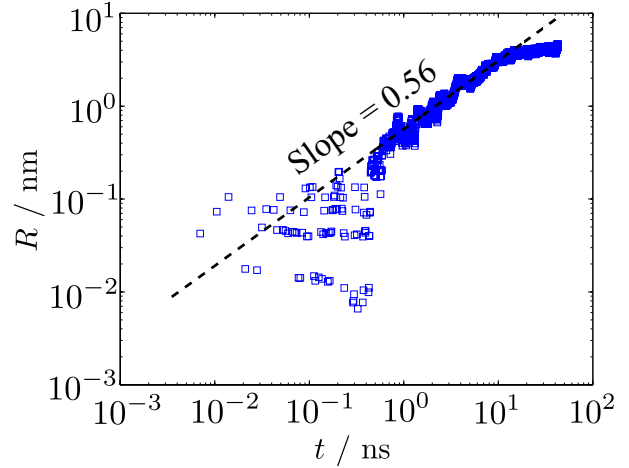}
  \caption{\label{fig:diffgrowth}Bubble radius as a function of time during growth. The data shown represents bubble VIII, as it grows towards its equilibrium size and has been shifted such that $R(t=0)=0$. At intermediate times an exponent of 0.56 is observed, consistent with diffusive bubble growth. In the final stage the availability of gas is too low to sustain the growth rate, and the bubble settles at its equilibrium radius.}
\end{center}
\end{figure}
The stage between nucleation of a gas bubble and it reaching its final size is governed by diffusive bubble growth.
The relation between $R$ and $t$ for diffusive bubble growth is well known, namely a square root power law~\cite{Brennen,Gor09},
\begin{equation}
\label{eq:diffgrowth}
R(t)\sim (D t)^{\frac{1}{2}}\;,
\end{equation}
where $D$ is the diffusion constant of the gas in the liquid.
In Fig.~\ref{fig:diffgrowth} we show the growth of bubble VIII on a log-log scale. 
The curve is shifted in time such that $R(t=0)=0$.
A power law behaviour with exponent 0.56 is observed.
This is very close to the expected exponent of $\frac{1}{2}$, suggesting that the growth of the gas bubbles in the simulation is indeed limited by diffusion.
The power law of Eq.~\eqref{eq:diffgrowth} is derived assuming that the magnitude of the Laplace pressure is small compared to atmospheric pressure. 
This condition is not fulfilled in particular for the tiny bubbles in the beginning, which presumably accounts for the observed deviation from Eq.~\eqref{eq:diffgrowth}.
Also, the final stage of bubble growth ($t\gtrsim 10$~ns) exhibits a different behaviour with time, which is caused by the closed nature of the simulation system and thus the limited amount of gas available.
In this stage the available gas in the system is depleted, and the bubble assumes its (final) equilibrium radius.

\subsection{Equilibrium radius}
Finally, the bubble reaches its equilibrium radius.
When a bubble with radius $R$ exists inside an infinite body of liquid, the concentration of dissolved gas just outside the bubble ($r=R$) is given by Henry's law:
\begin{equation}
\label{eq:Henry}
\frac{p}{\phi} = k_H\;.
\end{equation}
Here, $p$ is the partial pressure of a specific gas in the gas phase, $\phi$ the gas concentration of that specific gas inside the liquid, and $k_H$ is Henry's constant. 
In this work we will use the (dimensionless) mole fraction of the gas and liquid as the concentration $\phi$.
The total pressure in a (2D) bubble is given by:
\begin{equation}
p_{b}=p_l+\frac{\gamma}{R}\;,
\end{equation}
where $p_l$ is the pressure in the surrounding liquid, which is usually negligible in the case of nanobubbles.
From this relation, we can see that surface tension is a strong driving force for dissolution, especially as the bubble radius $R$ becomes small.

$k_H$ is Henry's constant and depends on the type of liquid and the type of gas involved. $k_H$ is temperature-dependent, but in this work we will consider a system with a fixed temperature.
In separate measurements (in a system consisting of a liquid phase in equilibrium with a gas phase at $p=1$~atm) we found that $k_H\sim 10^9$~Pa for the Lennard-Jones fluids considered in this work.

A gas concentration gradient induces a diffusive mass flux $J$ according to Fick's law:
\begin{equation}
J=-D \vec{\nabla} \phi\;.
\end{equation}
In the case of a cluster of bubbles, there is a limited amount of liquid present between neighbouring bubbles.
Since the amount is limited (and small if the bubbles are sufficiently close to each other) the concentration is noticeably affected by the gas flow out of the bubbles.
Eventually, as more and more gas enters this space, the gas concentration at the mid-point between these bubbles will reach the gas concentration at $R$ as prescribed by Henry's law [Eq.~\eqref{eq:Henry}].
When this has happened, there is no concentration gradient any more (the concentration is equal everywhere) and no diffusive gas flow exists: the bubbles are stable.

Using Henry's law we can predict the equilibrium size of these nanobubbles.
For this, we have to express the concentration of the gas in the liquid as a function of the bubble radius $R$.
First, we know that the amount of gas in the liquid is simply given by:
\begin{equation}
\label{eq:ngl}
n_g^l(R) = N_g - n_g^b(R)\;,
\end{equation}
where $n_g^l$ is the number of gas molecules dissolved in the liquid, $N_g$ the total number of gas atoms in the system, and $n_g^b$ the amount of gas molecules inside the bubble.
The amount of gas inside the bubble can be related to the size of the bubble, using the Laplace pressure and the ideal gas law:
\begin{equation}
\label{eq:ngb}
n_g^b(R) = \frac{2\pi \gamma R d}{k_BT}\;.
\end{equation}
Here, $k_B$ is Boltzmann's constant, $T$ the temperature of the system.
To obtain this relation, we assumed that the Laplace pressure difference is much greater than the ambient pressure, which is a valid assumption for nanobubbles exposed to atmospheric ambient pressure.
Henry's law [Eq.~\eqref{eq:Henry}] dictates the equilibrium condition, so by combining Eqs.~\eqref{eq:ngl} and \eqref{eq:ngb} we can now solve for the equilibrium bubble radius $R_{eq}$:
\begin{equation}
\label{eq:Reqveel}
R_{eq}=\frac{k_BTN_g}{4\pi\gamma d}\left(1\pm\sqrt{1-4\frac{\pi\gamma^2 d N_l}{k_BTk_H N_g^2}}\right)\;.
\end{equation}
Since the equation is quadratic we get two solutions, of which we cannot say \emph{a priori} which one is valid, because both are positive and finite.


\subsection{Comparison with simulations}

\begin{figure}[ht!]
\begin{center}
  \includegraphics[width=80mm]{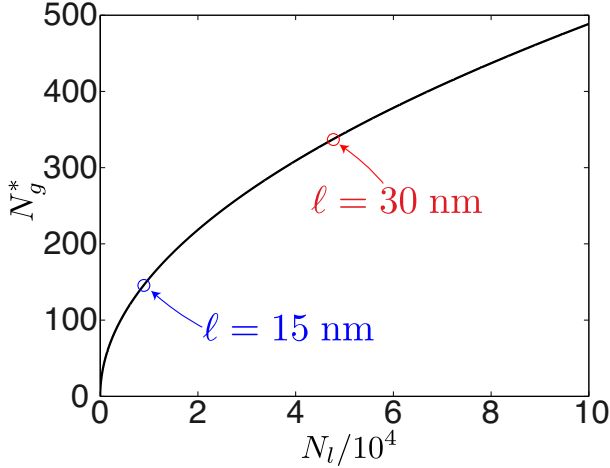}
  \caption{\label{fig:critgas}Critical number of gas atoms $N_g^\ast$ against $N_l$. The critical number of gas atoms is the amount of gas atoms required to prevent full dissolution of a nanobubble. As the system is larger (hence larger separation distance between nanobubbles) more gas is required to sustain a stable nanobubble.}
\end{center}
\end{figure}

We can now compare the simulation results to the model [Eq.~\eqref{eq:Reqveel}].

From Eq.~\eqref{eq:Reqveel} it is clear that for a physical solution to exist the number of gas atoms $N_g$ must be larger than:
\begin{equation}
\label{eq:Ngcrit}
N_g^\ast := \sqrt{\frac{4\pi\gamma^2 d N_l}{k_BTk_H}}\;.
\end{equation}
Hence, $N_g^\ast$ is a critical amount of gas atoms: When fewer gas atoms are present in the system the bubble will completely dissolve.

In Fig.~\ref{fig:critgas} this critical gas content as a function of $N_l$, which is proportional to the system size is shown. 
In the figure the two system sizes considered in this study are annotated.
From the figure it is clearly that for larger systems (hence, larger nanobubble seperations and higher $N_l$) the required amount of gas increases.
Intuitively, one would expect that the ratio $N_g^\ast/N_l$ is constant, but from Eq.~\eqref{eq:Ngcrit} and Fig.~\ref{fig:critgas} we can see that this is not the case.
Instead, the minimum number of gas molecules scales with the square root of the amount of liquid present, meaning that it scales linearly with the interbubble distance $\ell$.

\begin{figure*}[ht]
\begin{center}
  \includegraphics[width=160mm]{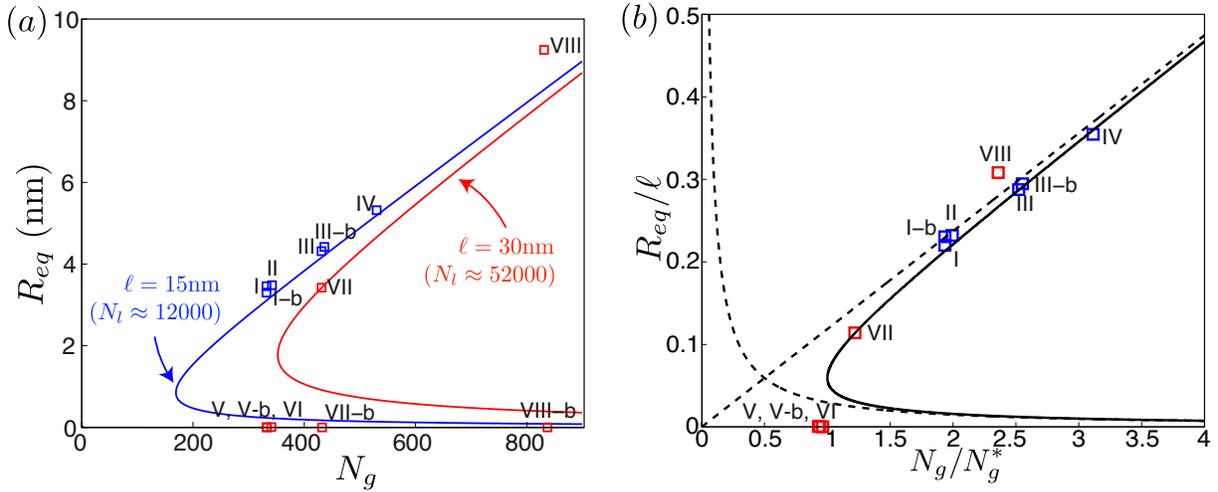}
  \caption{\label{fig:compare}(\emph{a}) Equilibrium radii $R_{eq}$ as measured in the simulations (symbols) and predicted by the model (lines) as a function of $N_g$. The blue data correspond to the small system ($\ell=15$nm), whereas the red data correspond to the large system ($\ell=30$nm). The model nicely predicts the equilibrium size in both systems, but of course does not account for the configurations where no bubble nucleation occurs (V-b, VII-b, and VIII-b). To obtain the theoretical curves, a value of $k_H=1.5\cdot 10^9$~Pa was used as a fit parameter. Independent measurements have shown $k_H\sim 10^9$~Pa. (\emph{b}) Collapse of the nondimensionalized equilibrium radii. All data points nicely follow the master curve. The simulations where no nucleation occurs are not included in this plot, to emphasize the good prediction of both $R_{eq}$ and $N_g^\ast$ by the model. The dashed lines indicate the expansion when $N_g\gg N_g^\ast$, Eq.~\eqref{eq:asympt}.}
\end{center}
\end{figure*}

The predicted equilibrium radius [Eq.~\eqref{eq:Reqveel}] as a function of gas content for the small systems ($\ell = 15$nm) and the large systems ($\ell = 30$ nm) are shown in Fig.~\ref{fig:compare}(a).
The measured equilibrium radii are also depicted, and show excellent agreement with the model.
For configurations V and VI (large systems), the model correctly predicts that there is no stable bubble size, hence they fully dissolve.
For configurations VII-b and VIII-b the model predicts that an equilibrium radius does exist, however in these cases there are no bubbles since the nucleation barrier is too high, which is not accounted for in the model.
By nondimensionalizing the equilibrium radius $R_{eq}$ with $\ell$, and rescaling $N_g$ with $N_g^\ast$ we can collapse the data on a single curve, independent of system size, see Fig.~\ref{fig:compare}(b).

For $N_g\gg N_g^\ast$, Eq.~\eqref{eq:Reqveel} can be expanded as:
\begin{eqnarray}
\label{eq:asympt}
\nonumber
R_{eq} = \frac{k_BTN_g}{4\pi\gamma d}\times \left[ 1\pm \sqrt{1-\left( \frac{N_g^\ast}{N_g}\right)^2} \right] = \\
\frac{k_BTN_g}{4\pi\gamma d} \times \left\{ \begin{array}{cc} 2+ \mathcal{O}\left(\left[ \frac{N_g^\ast}{N_g} \right] ^2\right) \\ \frac{1}{2}\left( \frac{N_g^\ast}{N_g} \right) ^2 + \mathcal{O}\left(\left[ \frac{N_g^\ast}{N_g} \right] ^4\right) \end{array} \right.  .
\end{eqnarray}
These asymptotic solutions are plotted in Fig.~\ref{fig:compare}b as the black dashed lines.
The first solution of Eq.~\eqref{eq:asympt} represents the limit where all gas atoms are contained within the bubble.
This can also be seen by comparing this result with Eq.~\eqref{eq:ngb}.
It is apparent from Fig.~\ref{fig:compare} that almost all bubbles are very close to this limit i.e. there is virtually no gas dissolved inside the liquids in the simulations.

\section{Nanobubble cluster stability}
\label{sec:cluststab}
Finally, we will discuss the stability of the entire cluster of nanobubbles.
Although it is now clear that bubbles that are surrounded by mirror images can indeed be in equilibrium, this does not mean that the entire cluster is stable.
There are two obvious threats to this stability: (\emph{i}) the Brownian motion of bubbles can cause neighboring bubbles to collide, leading to coalescence and coarsening of bubbles within the cluster and (\emph{ii}) when neighboring bubbles are not exactly equally sized, smaller bubbles would drain into larger bubbles via diffusion (similar to Ostwald ripening).

First, we address the possibility of bubbles colliding.
Since the distance between neighbouring bubbles is at best ten times as large as the bubble radius, Brownian motion of nanobubbles can indeed lead to collisions (see Fig.~\ref{fig:snaps}, where the nanobubbles are shown to move around).
An obvious way to prevent this from happening is to make sure that the bubbles repel each other.
This repulsion could be electrostatic (e.g. by using an ionic surfactant, or by the intrinsic negative charge of air bubbles in water \cite{Graciaa95,Takahashi05,Creux07}), which is sufficiently long-ranged.
Adding salt to the solution, which leads to screening of the electrostatic fields around the nanobubbles, would then reduce the stability of bulk nanobubble clusters which has indeed been observed in experiments~\cite{Jin07}.

The second issue (Ostwald ripening) is harder to prevent, and we expect that in time coarsening would indeed occur.
Of course, the larger bubbles that are formed in this way are still able to provide some shielding, but the polydispersity of bubble sizes will make a theoretical analysis difficult.
Therefore, at this point, we do not have an explanation for the measured stability by Ohgaki~\emph{et al.}~\cite{Ohgaki10} where Ostwald ripening apparently has been suppressed.

\section{Conclusions}
In conclusion, we have shown that bulk nanobubble clusters can indeed be stable under specific conditions.
First, the individual bubbles are surrounded by similar nanobubbles.
The distance between these bubbles must be small enough, such that the bubbles can succeed in saturating the liquid between the bubbles with gas, before the bubbles are completely drained.
When a cluster of stable bubbles exist, and the system is closed (so no gas can escape) the bubbles will in principle live forever, as long as bubbles cannot merge and Ostwald ripening is somehow prevented.
The merging of bubbles can be prevented by ionic surfactants.
Ostwald ripening is harder to prevent in theory, but experimental results showing that nanobubble cluster can indeed be stable for longer times indicate that there exists a mechanism that can prevent Ostwald ripening from occuring~\cite{Ohgaki10,Jin07,Jin07_2}.


\section*{Acknowledgements}
This research was stimulated by the organisers and participants of the ``Nanobubbles in Biology" workshop, UK, 2011.
This work was sponsored by the NCF for the use of supercomputer facilities and FOM, both with financial support from the Nederlandse Organisatie voor Wetenschappelijk Onderzoek (Netherlands Organisation for Scientific Research, NWO).

\end{document}